\documentstyle[aps]{revtex}
\begin{document}

\begin{center}
{\Large Spin-Peierls Dimerization of a $s=\frac 12$ Heisenberg
Antiferromagnet on a Square Lattice}\\[0pt]
\vspace{1cm}

Aiman Al-Omari\cite{mail}\\[0pt]

Department of Physics, \\[0pt]
Quaid-i-Azam University, \\[0pt]
Islamabad, Pakistan 45320.\\[0pt]

\vspace{1cm}(October 1998)\vspace{1.5cm}

{\bf Abstract}\vspace{0.7cm}
\end{center}

Dimerization of a spin-half Heisenberg antiferromagnet on a square lattice
is investigated for several possible dimerized configurations, some of which
are shown to have lower ground state energies than the others. In
particular, the lattice deformations resulting in alternate stronger and
weaker couplings along both the principal axes of a square lattice are shown
to result in a larger gain in magnetic energy. In addition, a `columnar'
configuration is shown to have a lower ground state energy and a faster
increase in the energy gap parameter than a `staggered' configuration. The
inclusion of unexpanded exchange coupling leads to a power law behaviour for
the magnetic energy gain and energy gap, which is qualitatively different
from that reported earlier. Instead of increasing as $\delta ^x$, the two
quantities depend on $\delta $ as $\delta ^\nu /\left| \ln \delta \right| .$
This is true both in the near critical regime $(0\leq \delta \leq 0.1)$ as
well as in the far regime $(0\leq \delta <1)$. It is suggested that the
unexpanded exchange coupling is as much a source of the logarithmic
dependence as a correction due to the contribution of umklapp processes.
Staggered magnetization is shown to follow the same $\delta $-dependence in
all the configurations in the small $\delta $-regime, while for $0\leq
\delta <1$, it follows the power law $\delta ^x$.\vspace{1cm}

\vspace{0.5in}

\ 

\section{Introduction}

It is known that dimerization lowers the ground state energy of a spin-half
isotropic Heisenberg antiferromagnet \cite
{soos,jacobs,cross,fields,matsuyama,bonner,soos1,spronken,guo,chitra}. In
other words, the system stands to gain energy by such lattice deformations
that render it dimerized with alternate weaker and stronger bonds between
spins on neighboring sites. On the other hand the lattice distortions cost
energy and it is the net energy balance that would determine whether the
gain in magnetic energy $\varepsilon (\delta )-\varepsilon (0)$ is large
enough to affect the spin-Peierls transition through dimerization. In a
phenomenological theory, this is usually seen in terms of an exponent
showing the dependence of magnetic and elastic energies on the dimerization
parameter $\delta ;$ where $0\leq \delta <1$. The parameter $\delta $
describes the extent of lattice deformation, i.e., it gives the displacement
of the $i$th atom through $u_i=\frac 12(-1)^i\delta $. The spin-dimer
formation is usually described by the Hamiltonian 
\begin{equation}
H=J\sum_i[1+(-1)^i\delta ]{\bf S}_i\cdot {\bf S}_{i+1}
\end{equation}
envisaging alternate stronger and weaker exchange bonds $J(1+\delta )$ and $%
J(1-\delta )$. These bonds can , in fact, be seen to result from the ansatz $%
J(a)=\frac Ja$\cite{comment}$.$ Thus when the distance between a pair of
spins decreases from $a$ to $a(1-\delta )$, the exchange coupling is taken
to change from $J$ to approximately $J(1+\delta )$.\newline

Since $0\leq \delta \leq 1$, and since elastic energies go typically as $%
\delta ^2$, therefore if the magnetic energy gain varies with $\delta $ with
an exponent less than 2 then in the limit $\delta \rightarrow 0$, the gain
would overwhelm the cost. In situations where the ground state is amenable
to dimerization, the spin-Peierls transition will be unconditional \cite
{guo,tang}. \newline

Such aspects as these have been studied extensively in Heisenberg
antiferromagnetic chains, as summarized in Table 1. This aspect has also
been revealed by experiments on quasi-one dimensional Heisenberg
antiferromagnet CuGeO$_3$ \cite{hase,nishi,fujita}.\newline

The situation in two-dimensions is a little more involved because{\bf \ }of
the possibility of frustration due to a competing antiferromagnetic second
neighbour interaction which can in principle destroy any LRO of the Neel
type as well as the possibility of dimerization. Much of the study of
two-dimensional Heisenberg antiferromagnet has therefore remained focused on
the destruction of order by frustration\cite{gelfand}. Moreover,{\bf \ }the
ground state of a Heisenberg antiferromagnet on a square lattice at zero
temperature is Neel-ordered and a critical value of spin-lattice coupling is
required for the gain in magnetic energy to affect a spin-Peierls transition 
\cite{tang,zhang,feiguin}. It is assumed below that the spin-lattice
coupling is above the threshold, allowing for dimerization of the lattice.
The spin configuration is expected to remain Neel-like under dimerization.
This is true in the absence of either frustration or quantum fluctuations
which lead to a melting of the Neel lattice\cite{richter}. \newline

The matter of frustration and quantum fluctuation aside, a simple
dimerization of a square lattice is interesting in its own right because the
lattice distortions can take place in more than one way, each one of the
possible configurations giving a different dependence of the ground state
energy on the dimerization parameter.\newline

Figure (1) shows a few such configurations. Fig.(a) describes a columnar
configuration caused by one longitudinal static ($\pi ,0$) phonon, in which
the nearest neighbour distances along the $x$-axis are taken to vary
alternately as $a(1+\delta )$, while those along the $y$-direction remain $a$%
. Fig.(b) shows a staggered configuration in which the lattice deformation
along the $x$-direction is alternated as in Fig.(a), but the sequence of
alternations is itself alternated as one goes along the $y$-direction. It is
caused by a $(\pi ,\pi )$ phonon with polarization along the $x$-axis\cite
{tang}. The difference between the earlier considerations of this
configuration and ours is that we take into account the elongation in the
exchange bond along the $y$-direction also, making it dependent upon the
dimerization parameter $\delta $. While the coupling along the $x$-direction
is alternately $\frac J{1-\delta }$ and $\frac J{1+\delta }$, it is
uniformly $\frac J{\sqrt{(1+\delta ^2)}}$ along the $y$-direction.\newline

In contrast to the configurations (a) and (b), those in Figs.(c), (d) and
(e) allow for simultaneous dimerization along both $x$- and $y$-directions
in the plane. The difference between (c) and (d) is the same as that between
(a) and (b): configuration (c) is columnar and (d) is staggered. The former,
caused by two phonons with wavevectors $(\pi ,0)$ and $(0,\pi )$, is called
plaquette configuration \cite{tang,leung}. Fig.(e) shows a much studied
configuration, caused by a longitudinal $(\pi ,\pi )$ phonon mode. In these
three configurations also the exchange couplings in both $x-$ and $y-$
directions are $\delta $-dependent. These five configurations of a dimerized
square lattice consisting of $N$ spins are therefore characterized by the
following nearest neighbour interactions.\newline

{\bf Configuration (a)}\newline

$J_{x,\lambda }=\frac J{(1+\lambda \delta )}\simeq J(1-\lambda \delta ),$ $%
\lambda =\pm 1$

$J_y=J.$\newline

That is to say, the dimerization is described by the Hamiltonian

\begin{eqnarray}
H=J\sum_{i,j}^{\sqrt{N}}\left[ \frac 1{\left( 1+(-1)^i\delta \right) }{\bf S}%
_{i,j}\cdot {\bf S}_{i+1,j}+{\bf S}_{i,j}\cdot {\bf S}_{i,j+1}\right]
\label{h1}
\end{eqnarray}

{\bf Configuration (b)}\newline

$J_{x,\lambda }=\frac J{(1+\lambda \delta )}\simeq J(1-\lambda \delta
),\lambda =\pm 1$

$J_y=\frac J{\sqrt{1+\delta ^2}}\simeq J(1-\frac{\delta ^2}2)$\newline

and the Hamiltonian is given by

\begin{eqnarray}
H=J\sum_{i,j}^{\sqrt{N}}\left[ \frac 1{\left( 1+(-1)^{i+j}\delta \right) }%
{\bf S}_{i,j}\cdot {\bf S}_{i+1,j}+\frac 1{\sqrt{1+\delta ^2}}\,\,{\bf S}%
_{i,j}\cdot {\bf S}_{i,j+1}\right] .  \label{h2}
\end{eqnarray}

{\bf Configuration (c)}\newline

$J_{x,\lambda }=J_{y,\lambda }=\frac J{(1+\lambda \delta )}\simeq
J(1-\lambda \delta ),$ $\lambda =\pm 1$ \newline

with the Hamiltonian

\begin{eqnarray}
H=J\sum_{i,j}^{\sqrt{N}}\left[ \frac 1{\left( 1+(-1)^i\delta \right) }{\bf S}%
_{i,j}\cdot {\bf S}_{i+1,j}+\frac 1{\left( 1+(-1)^j\delta \right) }{\bf S}%
_{i,j}\cdot {\bf S}_{i,j+1}\right] .  \label{h3}
\end{eqnarray}

{\bf Configuration (d)}\newline

$J_{x,\lambda }=\frac J{(1+\lambda \delta )}\simeq J(1-\lambda \delta ),$ $%
\lambda =\pm 1$\newline

$J_{y,\lambda }=\frac J{\sqrt{\delta ^2+\left( 1+\lambda \delta \right) ^2}}%
\simeq J\left( 1-\lambda \delta -(1-\frac{\lambda ^2}2)\delta ^2\right) $%
\newline

and the Hamiltonian

\begin{equation}
H=J\sum_{i,j}^{\sqrt{N}}\left[ \frac 1{\left( 1+(-1)^{i+j}\delta \right) }%
{\bf S}_{i,j}\cdot {\bf S}_{i+1,j}+\frac 1{\sqrt{\delta ^2+\left(
1+(-1)^j\delta \right) ^2}}\,\,{\bf S}_{i,j}\cdot {\bf S}_{i,j+1}\right] .
\label{h4}
\end{equation}

{\bf Configuration (e)}\newline

$J_{x,\lambda }=J_{y,\lambda }=\frac J{\sqrt{\delta ^2+\left( 1+\lambda
\delta \right) ^2}}\simeq J\left( 1-\lambda \delta -(1-\frac{\lambda ^2}2%
)\delta ^2\right) ,$ $\lambda =\pm 1$\newline

and the Hamiltonian 
\begin{equation}
H=J\sum_{i,j}^{\sqrt{N}}\frac 1{\sqrt{\delta ^2+\left( 1+(-1)^{i+j}\delta
\right) ^2}}\left[ {\bf S}_{i,j}\cdot {\bf S}_{i+1,j}+\,\,{\bf S}_{i,j}\cdot 
{\bf S}_{i,j+1}\right] .  \label{h5}
\end{equation}

Some of the exchange couplings in Eqs.(\ref{h1}-\ref{h5}) blow up at $\delta
=1$. Our analysis will therefore be confined to $0\leq \delta <1$.\newline

We would like to investigate the five configurations in order to see (i)
which of them gives the largest gain in magnetic energy as the dimerization
sets in, and (ii) whether the use of untruncated exchange coupling leads to
a single power law valid for the entire range of $\delta $.\newline

A number of methods can be chosen for this purpose. Spin wave theory, either
modified through Takahashi constraint of zero magnetization or a
Hartree-Fock approximated non-linear theory, is known to give surprisingly
good results for spin-half Heisenberg antiferromagnet. Or, a spin wave
theory in the spinless fermionic representation through Jordan-Wigner
transformations takes care of fermionic correlations among the $s=\frac 12$
spins. Coupled cluster method (CCM) has also been extensively, and
successfully, used for spin-half Heisenberg antiferromagnet in one and two
space dimensions.\newline

The first two methods belong to the class of mean field theories and hence
are not expected to be very reliable when it comes to determining critical
exponents. The coupled cluster method, on the other hand, is a perturbation
method in which increasingly higher order correlations can, in principle, be
incorporated at will, and which has been shown to give satisfactory results
even in the lower orders of perturbation\cite{bishop,bishop1}.{\bf \ }We
believe that the coupled cluster method must be sufficiently good to see
which of the alternative configurations proposed here is favoured once a
spin-Peierls transition sets in.\newline

\section{Application of the coupled-cluster method}

In the coupled cluster method it is first necessary to define a ket state
starting from a model state $\mid \phi >,$ which in our case is the Neel
state. The exact ground state $\mid \Psi >$ of the system can then be
postulated as 
\begin{equation}
\mid \Psi >=e^{{\cal S}}\mid \phi >  \label{bra}
\end{equation}
where ${\cal S}$ is the correlation operator defined for an $N$ particle
system as 
\begin{eqnarray}
{\cal S} &=&\sum_n\,{\cal S}_n\,\, \\
\mbox{with     }{\cal S}_n &=&\sum_{i_1...i_n}{\sf S}%
_{i_1,....,i_n}C_{i_1}^{\dagger }C_{i_2}^{\dagger }\cdot \cdot \cdot \cdot
\cdot \cdot C_{i_n}^{\dagger }
\end{eqnarray}
and $C_i^{\dagger }$ is the creation operator defined with respect to the
model state. The ground state energy can then be found as the eigenvalue of
the Hamiltonian in the proposed ground state\newline

$He^{{\cal S}}\mid \phi >=E_ge^{{\cal S}}\mid \phi >$.\newline

Taking inner product with $<\phi \mid e^{-{\cal S}}$ gives\newline

$E_g=<\phi \mid e^{-{\cal S}}He^{{\cal S}}\mid \phi >.$\newline

The product $e^{-{\cal S}}He^{{\cal S}}$ can be written as a series of
nested commutators in the well-known expansion 
\begin{equation}
e^{-{\cal S}}He^{{\cal S}}=H+[H,{\cal S}]+\frac 1{2!}\left[ [H,{\cal S}],%
{\cal S}\right] +\cdot \cdot \cdot \cdot \cdot \cdot \cdot \cdot \cdot
\end{equation}
where in the present case the series terminates after the fourth term.%
\newline

It is usually easier to deal with the $s=\frac 12$ Heisenberg Hamiltonian by
applying a rotation of 180$^{\circ }$ to the up spin sublattice; $%
S_x\rightarrow -S_x,$ $S_y\rightarrow S_y$ and $S_z\rightarrow -S_z$ such
that all the spins in the lattice point down. It is also convenient to
replace the spin operators with Pauli matrices: $S^j=\frac 12\sigma ^j,$ $%
j=x,y,z$ \cite{bishop,bishop1}. A general expression for the nearest
neighbour spin Hamiltonian in 2D is then 
\begin{equation}
H=-\frac J4\sum_{{\bf l,\rho }}\left( 2(\sigma _{{\bf l}}^{+}\sigma _{{\bf %
l+\rho }}^{+}+\sigma _{{\bf l}}^{-}\sigma _{{\bf l+\rho }}^{-})+\sigma _{%
{\bf l}}^z\sigma _{{\bf l+\rho }}^z\right) ,
\end{equation}
where ${\bf \rho }$ is a vector to the four nearest neighbours.
Correspondingly, the string operator ${\cal S}_n$ can now be defined as 
\begin{equation}
{\cal S}_{2n}=\frac 1{(n!)^2}\sum_{{\bf i}_1{\bf ....i}_n}\sum_{{\bf j}_1%
{\bf ....j}_n}{\sf S}_{{\bf i}_1{\bf ...i}_n;\,\,{\bf \,j}_1{\bf ....j}%
_n}\sigma _{{\bf i}_1}^{+}\sigma _{{\bf i}_2}^{+}\cdot \cdot \cdot \sigma _{%
{\bf i}_n}^{+}\sigma _{{\bf j}_1}^{+}\sigma _{{\bf j}_2}^{+}\cdot \cdot
\cdot \sigma _{{\bf j}_n}^{+},
\end{equation}

where subscripts ${\bf i}$ and ${\bf j}$ distinguish between sites on the
two sublattices. We note that for spin half $(\sigma _{{\bf l}%
}^{+})^2=(\sigma _{{\bf l}}^{-})^2=0.$ Truncation of the summation up to the
desired level gives rise to different schemes of approximation. Taking
interaction only between the spins on adjacent sites gives the so-called SUB$%
_{2-2}$ scheme. Including interactions with the second and fourth
neighboring sites gives what is termed as SUB$_{2-4}$ scheme. And taking the
previous two schemes including interaction among the four adjacent sites
give us what has been termed as local SUB$_4$, or LSUB$_4$ for short. Each
one of these approximations accounts for a different order of perturbation
calculation, and takes into account a different order of inter-particle
correlations. It has been noted that LSUB$_4$ is a sufficiently good
approximation for calculating the ground state properties of a spin-half
Heisenberg system \cite{bishop1}.\newline

Consider a general case: a Hamiltonian which has four different coupling
constants for nearest neighbour interactions in two space dimensions. It can
be written as

\begin{eqnarray}
H=-\frac 14\sum_{i,j}^{\sqrt{N}/2}\sum_{\lambda =\pm 1}\left[ J_{x,\lambda
}\,\,\,{\bf \sigma }_{2i,j}\cdot {\bf \sigma }_{2i+\lambda ,j}+J_{y,\lambda
}\,\,\,{\bf \sigma }_{i,2j}\cdot {\bf \sigma }_{i,2j+\lambda }\right] .
\end{eqnarray}

Here $i$ and $j$ are the two components of the site indices on a square
lattice. The correlation operators in the LSUB$_4$ scheme are defined as%
\newline
\begin{eqnarray}
{\cal S}_2 &=&\sum_{i,j}\left[ a_1\,\,\,\sigma _{2i,j}^{+}\sigma
_{2i+1,j}^{+}+b_1\sigma _{2i,j}^{+}\sigma _{2i-1,j}^{+}\,\,\,+c_1\sigma
_{i,2j}^{+}\sigma _{i,2j+1}^{+}+d_1\sigma _{i,2j}^{+}\sigma
_{i,2j-1}^{+}\right]  \nonumber \\
{\cal S}_3 &=&\sum_{i,j}\left[ a_3\sigma _{2i,j}^{+}\sigma
_{2i+3,j}^{+}\,\,\,+b_3\sigma _{2i,j}^{+}\sigma
_{2i-3,j}^{+}\,\,\,+c_3\sigma _{i,2j}^{+}\sigma _{i,2j+3}^{+}+d_3\sigma
_{i,2j}^{+}\sigma _{i,2j-3}^{+}\right]  \nonumber \\
{\cal S}_4 &=&\sum_{i,j}\left[ f\,\,\prod_{\nu =0}^3\sigma _{2i+\nu
,j}^{+}+g\,\prod_{\nu =0}^3\sigma _{2i-\nu ,j}^{+}+h\,\prod_{\nu =0}^3\sigma
_{i,2j+\nu }^{+}\,+l\,\,\prod_{\nu =0}^3\sigma _{i,2j-\nu }^{+}\right]
\end{eqnarray}

In these equations, the coefficients $a_{1,}b_{1,}$etc., are various forms
of the coefficient {\sf S}$_{{\bf i}_1{\bf ...i}_n\,;\,\,{\bf j}_1{\bf ....j}%
_n}$ in the expressions for ${\cal S}_{2n}$. The ground state energy within
the LSUB$_4$ approximation comes out to be

\begin{equation}
\text{ }E_g=-\frac 1{16}\left[ J_{x,+1}\,\,\,\left( 1+4a_1\right)
+J_{x,-1}\,\,\,(1+4b_1)+J_{y,+1}\,\,\,(1+4c_1)+J_{y,-1}\,\,\,(1+4d_1)\right]
\end{equation}

The coefficients $a_1,a_2,\cdot \cdot \cdot ,l$ are obtained as solutions of
a set of coupled nonlinear equations. These equations arise from the fact
that such matrix elements as $<\phi \mid {\bf O}$ $e^{-{\cal S}}He^{{\cal S}%
}\mid \phi >$ are all zero when the operator ${\bf O}$ is any product of
creation operators, particularly if it is one of the operator products in
the correlation operator ${\cal S}$ above.

\begin{eqnarray}
\ &<&\sigma _{i,2j}^{-}\sigma _{i,2j+\nu }^{-}e^{-{\cal S}}He^{{\cal S}}>=0;%
\hspace{3.5cm}\nu =\pm 1,\pm 3  \nonumber \\
\ &<&\sigma _{2i,j}^{-}\sigma _{2i+\nu ,j}^{-}e^{-{\cal S}}He^{{\cal S}}>=0;%
\hspace{3.5cm}\nu =\pm 1,\pm 3  \nonumber \\
\ &<&\sigma _{2i,j}^{-}\sigma _{2i+\nu ,j}^{-}\sigma _{2i+2\nu ,j}^{-}\sigma
_{2i+3\nu ,j}^{-}e^{-{\cal S}}He^{{\cal S}}>=0;\hspace{1cm}\nu =\pm 1 
\nonumber \\
\ &<&\sigma _{i,2j}^{-}\sigma _{i,2j+\nu }^{-}\sigma _{i,2j+2\nu }^{-}\sigma
_{i,j+3\nu }^{-}e^{-{\cal S}}He^{{\cal S}}>=0;\hspace{1cm}\nu =\pm 1
\end{eqnarray}
where ${\cal S}={\cal S}_2+{\cal S}_3+{\cal S}_4$. These equations translate
into the following twelve equations for the unknown parameters:\newline

$(a_1^2-1-2a_3b_1-2f)J_{x,+1}+(2a_1+2a_1b_1-2a_3b_3)J_{x,-1}{}=0$\newline

$(2b_1+2a_1b_1-2a_3b_3)J_{x,+1}+(b_1^2-1-2a_1b_3-2g)J_{x,-1}=0$\newline

${(2}a_3+2a_1a_3{)}J_{x,+1}{+(2}a_3-a_1^2+2a_3b_1-f{)}J_{x,-1}{=0}$ $\newline
$

$(2b_3-b_1^2+2a_1b_3-g)J{_{x,+1}+}(2b_3+2b_1b_3{)}J{_{x,-1}=0}$ $\newline
$

$(-2a_3b_1+2a_1f+a_3b_1^2-a_1a_3b_1{)}J{_{x,+1}}{+(}%
f-a_1^2+2b_1f+a_3g+2a_1a_3b_3{)}J{_{x,-1}=0}\newline
$

$%
(g-b_1^2+2a_1g+b_3f+2a_3b_1b_3)J_{x,+1}+(-2a_1b_3+2b_1g+a_1^2b_3-a_1b_1b_3)J_{x,-1}=0 
$ \newline

$(c_1^2-1-2c_3d_1-2h)J_{y,+1}+(2c_1+2c_1d_1-2c_3d_3)J_{y,-1}=0$ \newline

$(2d_1+2c_1d_1-2c_3d_3)J_{y,+1}+(d_1^2-1-2c_1d_3-2l)J_{y,-1}=0$\newline

$(2c_3+2c_1c_3)J_{y,+1}+(2c_3-c_1^2+2c_3d_1-h)J_{y,-1}=0 $ \newline

$(2d_3-d_1^2+2c_1d_3-l)J_{y,+1}+(2d_3+2d_1d_3)J_{y,-1}=0 $ \newline

$(-2c_3d_1+2c_1h+c_3d_1^2-c_1c_3d_1)J_{y,+1}+
(h-c_1^2+2d_1h+c_3l+2c_1c_3d_3)J_{y,-1}=0 $ \newline

$(l-d_1^2+2c_1l+d_3h+2c_3d_1d_3)J_{y,+1}
+(-2c_1d_3+2d_1l+c_1^2d_3-c_1d_1d_3)J_{y,-1}=0$\newline

Setting all the coupling constants $J_\mu $ equal reduces the number of
equations from twelve to three and yields exactly the same equations as
obtained by others \cite{bishop,bishop1}. The two sets of six equations each
independently determines the six coefficients contained in each of them. As
expected, the equations are symmetric in some coefficients. The twelve
coefficients are to be evaluated by solving the above coupled equations
numerically for each of the configurations separately by substituting
appropriate values of $J_{x,\lambda }$ and $J_{y,\lambda }$. \newline

To be able to calculate the energy gap between the ground and the first
excited states, we shall construct the excited ket state $|\Psi _e>$ in term
of a linear excitation operator ${\bf X},$ which, operating on the ground
state $|\Psi _0>,$ takes the system to an excited state: $|\Psi _e>={\bf X}$ 
$|\Psi _0>={\bf X}e^{{\cal S}}$ $|\phi >.$ This operator is constructed as a
linear combination of products of creation operators \cite{bishop1} 
\begin{equation}
{\bf X}=\sum_nX_n
\end{equation}
with 
\begin{equation}
X_n=\sum_{{\bf j}_1{\bf ....j}_n}{\bf \chi }_{\,\,{\bf \,j}_1{\bf ....j}%
_n}\sigma _{{\bf j}_1}^{+}\sigma _{{\bf j}_2}^{+}\cdot \cdot \sigma _{{\bf j}%
_n}^{+}.
\end{equation}
The first excited state is obtained by the operator 
\begin{equation}
X_1=\sum_{{\bf j}}{\bf \chi }_{\,\,\,{\bf j}}\sigma _{{\bf j}}^{+}
\end{equation}

where ${\bf j}$ can be any site of the two sublattices. It is easily seen
that the first excitation energy is 
\begin{equation}
E_e=\frac 18\left( \frac 12+2a_1+2b_1+2a_3+2b_3\right) \left(
J_{x,+1}+J_{x,-1}\right) .
\end{equation}
The energy gap for a given $\delta $ is $\Delta (\delta )=E_e(\delta
)-\left| E_g(\delta )\right| .$ We define gap parameter as $D(\delta
)=\Delta (\delta )-\Delta (0).$ This is the energy required to break a
dimerized singlet pair for a given $\delta $. \newline

To be able to calculate a quantity like staggered magnetization the need of
defining the bra state arises. In fact the bra state is not simply a
conjugate of the ket state defined in Eq.(\ref{bra}). The bra ground state
wave function $<\widetilde{\Psi }\mid $ corresponding to the ket state $%
|\Psi >$\ can be defined as\cite{bishop,bishop1} 
\[
<\widetilde{\Psi }\mid =<\phi \mid \widetilde{{\cal S}}_{2n}e^{-{\cal S}} 
\]
where the correlation operator $\widetilde{{\cal S}}$\ is built wholly out
of destruction operators of the Hamiltonian used. The need of defining such
operators comes from the fact that $e^{-{\cal S}}$\ is not equal to $e^{%
{\cal S}^{\dagger }}$. In our case the correlation operator is defines as%
{\bf \ } 
\[
\widetilde{{\cal S}}_{2n}=1+\sum_{n=1}^{N/2}\widetilde{s}_{2n} 
\]
with 
\[
\widetilde{s}_{2n}=\frac 1{(n!)^2}\sum_{{\bf i}_1{\bf ....i}_n}\sum_{{\bf j}%
_1{\bf ....j}_n}\widetilde{{\sf S}}_{{\bf i}_1{\bf ...i}_n;\,\,{\bf \,j}_1%
{\bf ....j}_n}\sigma _{{\bf i}_1}^{-}\sigma _{{\bf i}_2}^{-}\cdot \cdot
\cdot \sigma _{{\bf i}_n}^{-}\sigma _{{\bf j}_1}^{-}\sigma _{{\bf j}%
_2}^{-}\cdot \cdot \cdot \sigma _{{\bf j}_n}^{-}, 
\]
where ${\bf i}$\ and ${\bf j}$\ indicate vectors in the two sublattices
respectively. The first term in $\widetilde{{\cal S}}_{2n}$\ ensures
orthonormality of the bra and ket state; i.e., $<\widetilde{\Psi }\mid \Psi
> $ = $<\phi \mid \Psi >$ = $<\phi \mid \phi >$ = 1. The bra state
coefficients are found by putting the matrix elements of the commutator of
the Hamiltonian with a string of creation operators in the states $%
\widetilde{\Psi }$ and $\Psi $\ to zero{\bf . }

\begin{equation}
<\phi \mid \widetilde{{\cal S}}e^{-{\cal S}}[H,\sigma _{{\bf i}_1}^{+}\sigma
_{{\bf i}_2}^{+}\cdot \cdot \cdot \sigma _{{\bf i}_n}^{+}\sigma _{{\bf j}%
_1}^{+}\sigma _{{\bf j}_2}^{+}\cdot \cdot \cdot \sigma _{{\bf j}_n}^{+}]e^{%
{\cal S}}|\phi >=0,\vspace{.3 cm}n=1,2,3,\cdot \cdot \cdot \text{ }
\label{mag}
\end{equation}
{\bf \ }Equations(\ref{mag}) form a set of coupled linear equations for the
bra coefficients $\widetilde{{\sf S}}$, with the ket state coefficients
already known. It is to be noted here that the series of nested commutators
in $e^{-{\cal S}}[H,\sigma _{{\bf i}_1}^{+}\cdot \cdot \cdot \sigma _{{\bf j}%
_n}^{+}]e^{{\cal S}}$\ terminates after a finite number of terms.\newline

\ The correlation operators in the LSUB$_4$\ scheme are: \newline
\begin{eqnarray}
\widetilde{{\cal S}}_2 &=&\sum_{i,j}\left[ \widetilde{a}_1\,\,\,\sigma
_{2i,j}^{-}\sigma _{2i+1,j}^{-}+\widetilde{b}_1\sigma _{2i,j}^{-}\sigma
_{2i-1,j}^{-}\,\,\,+\widetilde{c}_1\sigma _{i,2j}^{-}\sigma _{i,2j+1}^{-}+%
\widetilde{d}_1\sigma _{i,2j}^{-}\sigma _{i,2j-1}^{-}\right]  \nonumber \\
\widetilde{{\cal S}}_3 &=&\sum_{i,j}\left[ \widetilde{a}_3\sigma
_{2i,j}^{-}\sigma _{2i+3,j}^{-}\,\,\,+\widetilde{b}_3\sigma
_{2i,j}^{-}\sigma _{2i-3,j}^{-}\,\,\,+\widetilde{c}_3\sigma
_{i,2j}^{-}\sigma _{i,2j+3}^{-}+\widetilde{d}_3\sigma _{i,2j}^{-}\sigma
_{i,2j-3}^{-}\right]  \nonumber \\
\widetilde{{\cal S}}_4 &=&\sum_{i,j}\left[ \widetilde{f}\,\,\prod_{\nu
=0}^3\sigma _{2i+\nu ,j}^{-}+\widetilde{g}\,\prod_{\nu =0}^3\sigma _{2i-\nu
,j}^{-}+\widetilde{h}\,\prod_{\nu =0}^3\sigma _{i,2j+\nu }^{-}\,+\widetilde{l%
}\,\,\prod_{\nu =0}^3\sigma _{i,2j-\nu }^{-}\right]
\end{eqnarray}
In these equations, the coefficients $\widetilde{a}_{1,}\widetilde{b}_{1,}$%
etc., are various forms of the coefficient $\widetilde{{\sf S}}_{{\bf i}_1%
{\bf ...i}_n\,;\,\,{\bf j}_1{\bf ....j}_n}$\ in the expressions for $%
\widetilde{s}_{2n}$. In the LSUB$_4$ scheme,\ the staggered magnetization,
given by{\bf \ } 
\[
M^z=-\frac 2N\sum_i<\sigma _i^z>, 
\]
where ${\bf i}$ runs over {\bf o}ne sublattice only, becomes 
\[
M^z=1-\widetilde{a}_1a_1-\widetilde{b}_1b_1-\widetilde{a}_3a_3-\widetilde{b}%
_3b_3-2\widetilde{f}f-2\widetilde{g}g-\widetilde{c}_1c_1-\widetilde{d}_1d_1-%
\widetilde{c}_3c_3-\widetilde{d}_3d_3-2\widetilde{h}h-2\widetilde{l}l 
\]

{\bf \ }

The bra state coefficients are determined from the following set of
simultaneous equations:\newline

{\bf \ }$(-\frac 12+\widetilde{b}_1b_1+\widetilde{a}_1a_1+\widetilde{a}_3a_3+%
\widetilde{b}_3b_3+b_1^2\widetilde{g}+b_1a_1\widetilde{f}+2\widetilde{g}g+2%
\widetilde{f}f)J_{x,+1}+$

\hspace{1.0in} $(\widetilde{a}_1+\widetilde{a}_1b_1-\widetilde{a}_3a_1-2%
\widetilde{f}a_1+2\widetilde{f}b_1a_1+\widetilde{f}a_3b_3+\frac 12\widetilde{%
g}b_3b_1)J_{x,-1}=0$\newline

$(\widetilde{b}_1+\widetilde{b}_1a_1-\widetilde{b}_3b_1-2\widetilde{g}b_1+2%
\widetilde{g}b_1a_1+\widetilde{g}b_3a_3+\frac 12\widetilde{f}%
a_1a_3)J_{x,+1}+ $

\hspace{1.0in}$(-\frac 12+\widetilde{a}_1a_1+\widetilde{b}_1b_1+\widetilde{a}%
_3a_3+\widetilde{b}_3b_3+\widetilde{f}a_1^2+\widetilde{g}b_1a_1+2\widetilde{f%
}f+2\widetilde{g}g)J_{x,-1}=0$\newline

$(-\widetilde{a}_1b_1-\widetilde{b}_1b_3+\widetilde{a}_3+\widetilde{a}_3a_1-2%
\widetilde{f}b_1+\widetilde{f}b_1^2-\widetilde{f}b_1a_1+\widetilde{g}%
b_1b_3)J_{x,+1}+$

\hspace{1.0in}$(-\widetilde{a}_1b_3+\widetilde{a}_3+\widetilde{a}_3b_1+%
\widetilde{f}b_3a_1+2\widetilde{f}g)J_{x,-1}=0$\newline

$(-\widetilde{b}_1a_3+\widetilde{b}_3+\widetilde{b}_3a_1+\widetilde{g}%
b_1a_3+2\widetilde{g}f)J_{x,+1}+$

\hspace{1.0in}$(-\widetilde{b}_1a_1-\widetilde{a}_1a_3+\widetilde{b}_3+%
\widetilde{b}_3b_1-2\widetilde{g}a_1+\widetilde{g}a_1^2-\widetilde{g}b_1a_1+%
\widetilde{f}a_1a_3)J_{x,-1}=0$\newline

$(-\widetilde{a}_1+2\widetilde{f}a_1+\widetilde{g}b_3)J_{x,+1}+(-\frac 12%
\widetilde{a}_1+\widetilde{f}+2\widetilde{f}b_1)J_{x,-1}=0$\newline

$(-\frac 12\widetilde{b}_1+\widetilde{g}+2\widetilde{g}a_1)J_{x,+1}+(-%
\widetilde{b}_1+2\widetilde{g}b_1+\widetilde{f}a_3)J_{x,-1}=0$\newline
\ 

$(-\frac 12+\widetilde{d}_1d_1+\widetilde{c}_1c_1+\widetilde{c}_3c_3+%
\widetilde{d}_3d_3+d_1^2\widetilde{l}+d_1c_1\widetilde{h}+2\widetilde{l}l+2%
\widetilde{h}h)J_{y,+1}+$

\hspace{1.0in}$(\widetilde{c}_1+\widetilde{c}_1d_1-\widetilde{c}_3c_1-2%
\widetilde{h}c_1+2\widetilde{h}d_1c_1+\widetilde{h}c_3d_3+\frac 12\widetilde{%
l}d_3d_1)J_{y,-1}=0$\newline

$(\widetilde{d}_1+\widetilde{d}_1c_1-\widetilde{d}_3d_1-2\widetilde{l}d_1+2%
\widetilde{l}d_1c_1+\widetilde{l}d_3c_3+\frac 12\widetilde{h}%
c_1c_3)J_{y,+1}+ $

\hspace{1.0in}$(-\frac 12+\widetilde{c}_1c_1+\widetilde{d}_1d_1+\widetilde{c}%
_3c_3+\widetilde{d}_3d_3+\widetilde{h}c_1^2+\widetilde{l}d_1c_1+2\widetilde{h%
}h+2\widetilde{l}l)J_{y,-1}=0$\newline

$(-\widetilde{c}_1d_1-\widetilde{d}_1d_3+\widetilde{c}_3+\widetilde{c}_3c_1-2%
\widetilde{h}d_1+\widetilde{h}d_1^2-\widetilde{h}d_1c_1+\widetilde{l}%
d_1d_3)J_{y,+1}+$

\hspace{1.0in}$(-\widetilde{c}_1d_3+\widetilde{c}_3+\widetilde{c}_3d_1+%
\widetilde{h}d_3c_1+2\widetilde{h}l)J_{y,-1}=0$\newline

$(-\widetilde{d}_1c_3+\widetilde{d}_3+\widetilde{d}_3c_1+\widetilde{l}%
d_1c_3+2\widetilde{l}h)J_{y,+1}+$

\hspace{1.0in}$(-\widetilde{d}_1c_1-\widetilde{c}_1c_3+\widetilde{d}_3+%
\widetilde{d}_3d_1-2\widetilde{l}c_1+\widetilde{l}c_1^2-\widetilde{l}d_1c_1+%
\widetilde{h}c_1c_3)J_{y,-1}=0$\newline

$(-\widetilde{c}_1+2\widetilde{h}c_1+\widetilde{l}d_3)J_{y,+1}+(-\frac 12%
\widetilde{c}_1+\widetilde{h}+2\widetilde{h}d_1)J_{y,-1}=0$\newline

$(-\frac 12\widetilde{d}_1+\widetilde{l}+2\widetilde{l}c_1)J_{y,+1}+(-%
\widetilde{d}_1+2\widetilde{l}d_1+\widetilde{h}c_3)J_{y,-1}=0$ \newline

\section{Results}

The ground state energy and magnetization can now be calculated as functions
of the dimerization parameter $\delta $ . Previous calculations have
invariably taken spin-spin exchange couplings alternately as $J(1\pm \delta
) $, which, as mentioned above, can be taken as an expansion of the
interaction in Eq.(\ref{h1}) to order $\delta $, implying that the results
are valid only in the critical regime $\delta \rightarrow 0$. We notice in
our calculations that if in Eqs.(\ref{h1}-\ref{h5}) all the expansions are
terminated at the order of $\delta $ then the distinction between
configurations (a) and (b) disappears. On the other hand, if the expansion
is taken to one order higher, then there remains no way to distinguish
between configurations (c) and (d). We must therefore either go to orders
beyond $\delta ^2$ in the expansion, or retain the interactions in their
unexpanded form. We do the latter. An added advantage is that the results
will then be valid in the limit $\delta \rightarrow 1$.\newline

\subsection{Magnetic energy gain}

Our calculations confirm that, like the chain, the ground state energy of
all the five configurations decreases with $\delta $. This is shown in
Figures 2, where $\varepsilon (\delta )-\varepsilon (0)$ is plotted against $%
\delta $ for the proposed configurations. This conclusion is not new for
some of the configurations in Fig.1\cite{tang,feiguin,rokhsar,read,xu,katoh}%
. However, what is significant is that the ground state energy goes down
with $\delta $ more rapidly for some configurations than others. In fact,
Fig. 2 shows that the $\delta $-dependence is markedly different for the two
types of dimerized configurations: one in which dimerization takes place
only along one axis, and the other, in which it occurs along both the
directions. The rate of decrease is significantly higher for the latter.
Also, the columnar configurations lead to a greater gain in magnetic energy
than the staggered ones. It also shows that the plaquette configuration of
Fig.1(c) is energetically the most favourable state, as also noted earlier 
\cite{tang,lieb}. Particularly in the complete range of $\delta $ ($0\leq
\delta <1$), the plaquette configuration stands out as the most preferred
one, while there is hardly a discernible difference among the configurations
(a), (b) and (d). \newline

Configuration (e) is peculiar in the sense that $\delta =\frac 12$ is a
special point for it; the shorter bond length is symmetric about this point,
having a minimum value of $\frac 1{\sqrt{2}}$. At this point the distortions
give rise to a rectangular lattice with sides $\sqrt{2}$ and $\frac 1{\sqrt{2%
}}$. The energy gain increases with $\delta $ up to $\delta =\frac 12$, and
then goes down.\newline

It is worth pointing out here that the much simpler mean field methods of
spin wave theory - either in the bosonic representation through
Holstein-Primakoff transformations, or in the fermionic representation
through Jordan-Wigner transformations - yield very similar results. This has
been checked by us separately.\newline

{\bf \ }Earlier calculations on{\bf \ }the spin-Peierls instability in a 2D
system gave varied results on the critical exponents. Monte Carlo
calculations of Tang and Hirsch \cite{tang} on the Hubbard model in the
limit of infinite on-site repulsion U found for the cases corresponding to
our configurations (a), (b), (c) and (e) that the magnetic energy gain
followed a simple power law behaviour and increased as $\delta ^2$. Their
cases are different from ours in the sense that couplings alternated as $%
J(1\pm \delta )$, and were taken constant along the $y$-direction in the
case (b). Feiguin et al \cite{feiguin} obtained similar results for
configurations (a) and (e) in the Schwinger boson representation. Quantum
Monte Carlo calculations of Katoh and Imada \cite{katoh} showed that in
chains that are coupled by an antiferromagnetic coupling the exponent of the
magnetic energy gain in the $\delta \rightarrow 0$ limit is 1.\newline

{\bf \ }Our results are expected to be different from these because instead
of $J(1\pm \delta )$, we take the unapproximated exchange coupling $J(a)=%
\frac Ja$. Our CCM calculations show that the gain in magnetic energy does
not vary with $\delta $ as a simple power law; it varies as $\frac{\delta ^v%
}{\left| \ln \delta \right| }$ for all the five configurations in the range $%
0\leq \delta \leq 0.1$ with the exponent $\nu =1.5$. In the complete range $%
0\leq \delta <1$ also, they show the same dependence on $\delta $ with $\nu
=1$ for the configurations (a) - (d).\newline

It is interesting to note that earlier results show, as summarized in Table
1, that the dimerization of an antiferromagnetic chain also varies as $\frac{%
\delta ^v}{\left| \ln \delta \right| }$, but only in the small $\delta $
regime (the near critical regime). There, the factor of $\frac 1{\left| \ln
\delta \right| }$ is brought about in the renormalization group calculations
as a correction due to umklapp processes \cite{soos1,black}. Our CCM
results, however, show that even in chains this may be the case when the
exchange couplings in the dimerized state are taken as $\frac J{1\pm \delta }
$, instead of the approximated $J(1\pm \delta )$. We find for chains that
the best fit is obtained with $\frac{\delta ^v}{\left| \ln \delta \right| }$
in the entire range of $\delta $ rather than only in the range of small $%
\delta $. With the full exchange couplings, the exponent for the chain comes
out to be $\nu =\frac 23$ for $0\leq \delta <1$, and $\nu =1.3$ - 1.6 for $%
0\leq \delta \leq 0.1$. The latter gives a decent comparison with the
numbers in Table 1.\newline

\subsection{The gap parameter}

The $\delta $ dependence of the energy gap parameter $D(\delta )$ defined
above for the five configurations is shown in Fig.3, showing greater
stabilization of the dimerized state with increasing $\delta $. We also find
that, like the magnetic energy gain, the gap parameter $D$ increases with $%
\delta $ as $\frac{\delta ^v}{\left| \ln \delta \right| }$ in the small $%
\delta $ regime for all the five configurations with $\nu =1.5$. The
configurations (a) - (d) also have the same dependence on $\delta $ in the
entire range of $\delta $ with $\nu =1$. \newline

The difference between the dimerization of a square lattice along only one
direction (Fig. 1(a) and (b)) and along both the directions (Fig. 1(c) ,(d)
and (e)) is again markedly brought out in Fig. 3. Also the columnar
configurations again appear as preferred modes of dimerization over the
staggered configurations for having higher values of the gap parameter in
the region of small $\delta $.\newline

\subsection{Staggered magnetization}

Our CCM calculations in the LSUB$_4${\bf \ }approximation give staggered
magnetization for the un-dimerized square lattice $M(\delta =0)$= 0.2965,
within about 2\% of the exact value of 0.303. As dimerization sets in,
magnetization decreases in all the configurations we have chosen, as shown
in Fig. 4, in agreement with the earlier results for configuration (a)\cite
{xu}. This is also the case for the entire range of $\delta $ ($0\leq \delta
<1$), except in the case of configuration (e) for which the magnetization
rises again after $\delta =\frac 1{\sqrt{2}}$.\newline

The CCM calculations show that for all the five configurations, the
magnetization also varies as $\frac{\delta ^v}{\left| \ln \delta \right| }$
in the small $\delta $ regime with the exponent $\nu =1.5$, exactly as the
energy gain and the gap parameter. However in the regime $0\leq \delta <1$, $%
M$ exhibits a simple power law dependence: $M$ $\symbol{126}$ $\delta ^x$
with $x$ between .65 and .75, as shown in Table 2. Configuration (e) has a
distinctly different behaviour in this regime.\newline

To summarize, we have studied the spin-Peierls dimerization of a spin-half
Heisenberg antiferromagnet on a square lattice taking unapproximated
exchange couplings based on the ansatz $J(a)=\frac{J}{a},$ and assuming that
the spin-lattice coupling is above the threshold to affect the spin-Peierls
transition. We have included different possibilities of dimerization. The
ground state energy as well as staggered magnetization decrease continuously
with increasing dimerization for all the proposed configurations. Of the
five configurations, those with dimerization taking place simultaneously
along both the principal square axes have markedly lower ground state
energies and magnetization than those with dimerization along only one of
the axes, in agreement with the result of Lieb and Nachtergaele \cite{lieb}.
Also, those with columnar dimerization have consistently lower energies than
those with the staggered dimerization. The plaquette configuration stands
out as the most favoured mode of dimerization. The energy gap parameter also
corroborates the above conclusions. It has also been shown that the magnetic
energy gain as well as the gap parameter and staggered magnetization depend
upon the dimerization parameter $\delta $ as $\frac{\delta ^{v}}{\left| \ln
\delta \right| }$, at least in the $\delta \rightarrow 0$ regime, the $%
\left| \ln \delta \right| $ factor coming in without any considerations of
umklapp processes being included.\

\ 

{\bf Figure captions}\newline

Figure 1: Five configurations for the dimerization of a square lattice. (a)
a columnar configuration caused by a longitudinal $(\pi ,0)$ static phonon.
The nearest neighbour coupling along the horizontal direction alternates
between $J(1-\delta )$ and $J(1+\delta )$, while that along the vertical
direction remains $J$. (b) a staggered configuration caused by a $(\pi ,\pi
) $ static phonon with polarization along{\bf \ }$x$-direction. Like (a),
the dimerization occurs along one direction only, but the sequence of
alternate couplings itself alternates along the other direction. The
coupling along the vertical direction is also taken to vary with $\delta $.
(c) Dimerization along both the directions, caused by $(\pi ,0)$ and $(0,\pi
)$ phonons, making a plaquette of four nearest neighbour spins. (d) Again
dimerization along both the directions, but taken staggered along the
vertical direction. {\bf \ }(e) Another staggered dimerization that is
caused by a longitudinal $(\pi ,\pi )$ phonon.{\bf \ }Chains are formed with
strong bonds. The square lattice deforms to a rectangular lattice with this
mode for $\delta =1/2$.\newline

Figure 2: The gain in magnetic energy $\varepsilon (\delta )-\varepsilon (0)$
as dimerization sets in with increasing $\delta $ for the five
configurations; (a) in the range $0\leq \delta \leq 0.1$, and (b) in $0\leq
\delta <1$.\newline

Figure 3: Dependence of the energy gap parameter $D$ on $\delta $ for the
five dimerization configurations; (a) in the range $0\leq \delta \leq 0.1$,
and (b) in $0\leq \delta <1$.\newline

Figure 4: Staggered magnetization varying with $\delta $ for the five
dimerization configurations; (a) in the range $0\leq \delta \leq 0.1$, and
(b) in $0\leq \delta \leq 1$.

\newpage\ 

Table-1 : Summary of the critical exponents for spin-Peierls transition in a
Heisenberg chain determined by various methods.\newline
\flushleft{\ }

\begin{tabular}{||c|c|c|c||}
\hline\hline
Method & Interval & $\varepsilon (\delta )-\varepsilon (0)$ & Exponent \\ 
\hline
{Random phase app.$^{{\cite{cross}}}$} & $0\leq \delta \leq 1$ & $\delta ^x$
& $x=4/3$ \\ \hline
{Renormalization group$^{{\cite{fields}}}$} & $0\leq \delta \leq 1$ & $%
\delta ^x$ & $x=1.53$ \\ \hline
{2-level RG$^{{\cite{matsuyama}}}$} & $0\leq \delta \leq 1$ & $\delta ^x$ & $%
x=1.78$ \\ 
& $0.05\leq \delta \leq 0.1$ & {$\delta ^{2\nu }$}${/}${$\mid \ln $}${(}${$%
\delta $}${)}${$\mid $} & $2\nu =1.68_{-0.36}^{+0.13}$ \\ 
& $\ 0.4\leq \delta \leq 0.5$ & {$\delta ^{2\nu }$}${/}${$\mid \ln $}${(}${$%
\delta $}${)}${$\mid $} & $2\nu =1.31\pm 0.02$ \\ \hline
{Excitation spectrum$^{{\cite{bonner}}}$} & $0\leq \delta \leq 1$ & $\delta
^x$ & $x=1.36_{-0.2}^{+0.1}$ \\ \hline
Valence bond$^{\cite{soos1}}$ & $\delta \leq 0.05$ & {$\delta ^{2\nu }$}${/}$%
{$\mid \ln $}${(}${\ $\delta $}${)}${$\mid $} & $\nu =2/3$ \\ 
& $\delta \geq 0.05$ & $\delta ^x$ & $x=1.36_{-0.2}^{+0.1}$ \\ \hline
{Finit size scaling$^{{\cite{spronken}}}$} & $0\leq \delta \leq 0.1$ & {$%
\delta $}${^{2\nu }/}${$\mid \ln $}${(}${$\delta $}${)}${$\mid $} & $\nu
=0.71\pm 0.01$ \\ 
& $0\leq \delta \leq 1$ & $\delta ^x$ & $x=1.34\pm 0.02$ \\ \hline
{Exact diagonalization$^{{\cite{guo}}}$} & $0\leq \delta \leq 0.1$ & ${%
\delta ^{2\nu }/\mid \ln (\delta )}${$\mid $} & $\nu =2/3$ \\ \hline
DMRG$^{\cite{chitra}}$ & \multicolumn{1}{|c|}{$\delta \leq 0.05$} & $\delta
^x$ & $x=2/3$ \\ \hline\hline
\end{tabular}

\vspace{1.0cm}\ \ 

{\ }\ Table 2: Exponents obtained by the CCM method for magnetic energy
gain, energy gap and magnetization for the five dimerized square lattice
configurations. The logarithmic power law goes as ${\delta ^\nu /\mid \ln
(\delta )\mid }$ in the five configurations for both energy gain and gap
parameter for both small and full $\delta $. While in the staggered
magnetization the logarithmic law is valid for $\delta \rightarrow 0$, but
it obeys a simple power law in the full dimerization limit; $\delta
\rightarrow 1$. Values of $\nu $ are listed in the following table.\newline

\flushleft{\ {\hspace{1in}}} 
\begin{tabular}{||c|c|c|c|c||}
\hline\hline
Configuration & Interval & $\varepsilon (\delta )-\varepsilon (0)$ & $\Delta
(\delta )-\Delta (0)$ & $M(0)-M(\delta )$ \\ \hline
(a) & $0\leq \delta \leq 0.1$ & $\nu =1.5$ & $\nu =1.5$ & $\nu =1.5$ \\ 
\cline{2-5}
& $0\leq \delta <1$ & $\nu =1.0$ & $\nu =1.0$ & $\sim {\delta ^{0.65}}$ \\ 
\hline
(b) & $0\leq \delta \leq 0.1$ & $\nu =1.5$ \thinspace \thinspace & $\nu =1.5 
$ & $\nu =1.5$ \\ \cline{2-5}
& $0\leq \delta <1$ & $\,\nu =1.0$ & $\nu =1.0$ & $\sim {\delta ^{0.75}}$ \\ 
\hline
(c) & $0\leq \delta \leq 0.1$ & $\nu =1.5$ & $\nu =1.5$ & $\nu =1.5$ \\ 
\cline{2-5}
& $0\leq \delta <1$ & $\nu =1.0$ & $\nu =1.0$ & $\sim {\delta ^{0.75}}$ \\ 
\hline
(d) & $0\leq \delta \leq 0.1$ & $\nu =1.5$ & $\nu =1.5$ & $\nu =1.5$ \\ 
\cline{2-5}
& $0\leq \delta <1$ & $\nu =1.0$ & $\nu =1.0$ & $\sim {\delta ^{0.65}}$ \\ 
\hline
(e) & $0\leq \delta \leq 0.1$ & $\nu =1.5$ & $\nu =1.5$ & $\nu =1.5$ \\ 
\cline{2-5}
& $0\leq \delta <1$ &  &  & $\sim {\delta ^{0.1}}$ \\ \hline\hline
\end{tabular}

\end{document}